\documentclass[aps,prl,twocolumn,groupedaddress]{revtex4}

\usepackage[dvips]{graphicx,color} 
\usepackage{array,hhline,dcolumn} 
\usepackage{rotating} 

\newcommand{\gtwid}{\mathrel{\raise.3ex\hbox{$>$\kern-.75em\lower1ex\hbox{$\sim$}}}}
\newcommand{\ltwid}{\mathrel{\raise.3ex\hbox{$<$\kern-.75em\lower1ex\hbox{$\sim$}}}}

\begin{document}
%

\title{A Search for Electron Antineutrino Appearance at the $\Delta m^2 \sim$ 1 $\mathrm{eV}^{2}$ Scale}

\author{
	A.~A. Aguilar-Arevalo$^{12}$, C.~E.~Anderson$^{15}$, S.~J.~Brice$^{5}$,
 	B.~C.~Brown$^{5}$, L.~Bugel$^{4}$, J.~M.~Conrad$^{11}$,
	Z.~Djurcic$^{4}$, B.~T.~Fleming$^{15}$, R.~Ford$^{5}$,
	F.~G.~Garcia$^{5}$, G.~T.~Garvey$^{9}$, J.~Gonzales$^{9}$,
	J.~Grange$^{6}$, C.~Green$^{5,9}$, J.~A.~Green$^{8,9}$,
	R.~Imlay$^{10}$,
	R.~A. ~Johnson$^{2}$, G.~Karagiorgi$^{11}$, T.~Katori$^{8,11}$,
	T.~Kobilarcik$^{5}$, S.~K.~Linden$^{15}$,
	W.~C.~Louis$^{9}$, K.~B.~M.~Mahn$^{4}$, W.~Marsh$^{5}$,
	C.~Mauger$^{9}$, V.~T.~McGary$^{11}$, 
	W.~Metcalf$^{10}$, G.~B.~Mills$^{9}$,
	C.~D.~Moore$^{5}$, J.~Mousseau$^{6}$, R.~H.~Nelson$^{3}$,
        P.~Nienaber$^{14}$, J.~A.~Nowak$^{10}$,
        B.~Osmanov$^{6}$, Z.~Pavlovic$^{9}$, D.~Perevalov$^{1}$,
	C.~C.~Polly$^{7,8}$, H.~Ray$^{6,9}$, B.~P.~Roe$^{13}$,
	A.~D.~Russell$^{5}$, M.~H.~Shaevitz$^{4}$, M.~Sorel$^{4}$\footnote{Present address: IFIC, Universidad de Valencia and CSIC, Valencia 46071, Spain},
	J.~Spitz$^{15}$, I.~Stancu$^{1}$, R.~J.~Stefanski$^{5}$,
	R.~Tayloe$^{8}$, M.~Tzanov$^{3}$, R.~G.~Van~de~Water$^{9}$, 
	M.~O.~Wascko$^{10}$\footnote{Present address: Imperial College; London SW7 2AZ, United Kingdom},
        D.~H.~White$^{9}$, M.~J.~Wilking$^{3}$, G.~P.~Zeller$^{4,9}$,
	E.~D.~Zimmerman$^{3}$ \\
\smallskip
(The MiniBooNE Collaboration)
\smallskip
}
\smallskip
\smallskip
\affiliation{
$^1$University of Alabama; Tuscaloosa, AL 35487 \\
$^2$University of Cincinnati; Cincinnati, OH 45221\\
$^3$University of Colorado; Boulder, CO 80309 \\
$^4$Columbia University; New York, NY 10027 \\
$^5$Fermi National Accelerator Laboratory; Batavia, IL 60510 \\
$^6$University of Florida; Gainesville, FL 32611 \\
$^7$University of Illinois; Urbana, IL 61801 \\
$^8$Indiana University; Bloomington, IN 47405 \\
$^9$Los Alamos National Laboratory; Los Alamos, NM 87545 \\
$^{10}$Louisiana State University; Baton Rouge, LA 70803 \\
$^{11}$Massachusetts Institute of Technology; Cambridge, MA 02139 \\
$^{12}$Instituto de Ciencias Nucleares, Universidad Nacional Aut\'onoma de M\'exico, D.F. 04510, M\'exico \\
$^{13}$University of Michigan; Ann Arbor, MI 48109 \\
$^{14}$Saint Mary's University of Minnesota; Winona, MN 55987 \\
$^{15}$Yale University; New Haven, CT 06520\\
}

\date{\today}

\begin{abstract}
The MiniBooNE Collaboration reports initial results from a search for 
$\bar{\nu}_{\mu}\rightarrow\bar{\nu}_e$ oscillations. A signal-blind analysis was performed
using a data sample corresponding to $3.39 \times 10^{20}$ protons 
on target. 
The data are consistent with background prediction across 
the full range of neutrino energy reconstructed
assuming quasielastic scattering, $200 < E_{\nu}^{QE} < 3000$ MeV: 144 electron-like
events have been observed in this energy range, compared to an expectation 
of $139.2 \pm 17.6$ events. No significant excess of events 
has been observed, both at low energy, 200-475 MeV, and at high energy, 
475-1250 MeV. The data are inconclusive with respect to antineutrino 
oscillations suggested by data from the Liquid Scintillator Neutrino Detector at Los Alamos National Laboratory.
\end{abstract}


\maketitle


Motivated by the LSND observation of an excess of $\bar{\nu}_e$ events in a $\bar{\nu}_{\mu}$ beam \cite{lsnd}, the MiniBooNE collaboration has previously performed a search for $\nu_\mu \rightarrow \nu_e$ oscillations, the results of which showed no evidence of an excess of $\nu_e$ events for neutrino energies above 475 MeV \cite{mb_osc,mb_lowe}. Assuming no CPT or CP violation, the results exclude the LSND excess interpreted as two-neutrino oscillations at $\Delta m^2 \sim$ 0.1-100 $\mathrm{eV}^{2}$ at 98\% C.L.. Similarly, the KARMEN experiment \cite{karmen} has performed a direct search for $\bar{\nu}_e$ appearance, and has placed a limit independent of any CPT or CP violation assumption. However, a joint analysis of KARMEN and LSND results shows high compatibility \cite{compatibility,mbcompatibility}. A corresponding $\bar{\nu}_{\mu}\rightarrow\bar{\nu}_e$ oscillation search has been performed at MiniBooNE and is presented in this Letter. This search serves as another direct test of LSND and provides complementary information to that of KARMEN, having sensitivity to the lower $\Delta m^{2}$ oscillations allowed by the joint KARMEN-LSND analysis \cite{nubarLOI}. It should be noted that, in a simple two-neutrino oscillation model where CPT conservation is imposed, oscillation probabilities (mixing amplitudes and mass-squared differences) for neutrinos and antineutrinos cannot be different. Therefore, the oscillation search presented here is a direct search for existence of non-standard oscillations where CPT is violated, or effectively violated.

Despite having observed no evidence for oscillations above 475 MeV, the MiniBooNE $\nu_{\mu}\rightarrow\nu_e$ search observed a 3.0$\sigma$ excess of electron-like events at low energy, between 200-475 MeV \cite{mb_lowe}. Although the excess is incompatible with LSND-type oscillations, several hypotheses \cite{sterile,hhh,weiler,goldman,nelson,kostelecky}, including sterile neutrino oscillations with CP violation, anomaly-mediated neutrino-photon coupling, and others, have been proposed that provide a possible explanation for the excess itself, and, in some cases, offer the possibility of reconciling the MiniBooNE $\nu_e$ excess with the LSND $\bar{\nu}_e$ excess. These phenomenological interpretations have provided additional motivation for an antineutrino appearance search at MiniBooNE. 

The analysis presented in this Letter mirrors the blind search performed in neutrino mode \cite{mb_osc}. It employs a two-neutrino oscillation model, where only $\bar{\nu}_{\mu}$ present in the MiniBooNE beam are allowed to oscillate into $\bar{\nu}_e$, at $\Delta m^{2} \sim$ 0.1-100 $\mathrm{eV}^2$. Given that no evidence of $\nu_{\mu}$ oscillations was observed in high-purity, high-statistics searches in neutrino mode \cite{mb_osc,mb_dis}, the analysis further assumes no $\bar{\nu}_{\mu}$ disappearance and no $\nu_{\mu}$ oscillations. In addition, no contribution from the observed neutrino mode low energy excess has been accounted for in the antineutrino prediction.

The antineutrino flux \cite{mb_flux} is produced by 8 GeV protons incident on a beryllium target. Negatively charged mesons produced in p-Be interactions are focused in the forward direction with the use of a toroidal magnetic field, and subsequently decay primarily into $\bar{\nu}_{\mu}$. In antineutrino mode, a large neutrino contamination ($\nu_{\mu}$ and $\nu_e$) of 15.9\% is expected in the flux viewed by the detector, compared to 5.9\% in neutrino mode. The intrinsic $\bar{\nu}_e$ and $\nu_e$ content is only 0.4\% and 0.2\%, respectively, coming primarily from $\pi\rightarrow\mu$ and $K$ decays. The $\bar{\nu}_{\mu}$ flux peaks at $\sim400$ MeV and has a mean energy of $\sim600$ MeV. See \cite{mb_flux} for more details.

A detailed description of the MiniBooNE detector is available in \cite{mb_detector}. The detector location was chosen to satisfy $L[\mathrm{m}]/E[\mathrm{MeV}]\sim1$, similar to that of LSND, thus maximizing sensitivity to oscillations at $\Delta m^2\sim1$ eV$^{2}$. The detector is filled with pure mineral oil (CH$_{2}$). Neutrino interactions in the detector produce final state electrons or muons, which produce scintillation and Cherenkov light detected by photomultiplier tubes (PMTs) that line the interior of the detector. The simulation of light incident on the PMTs takes into account decays and strong and electromagnetic re-interactions in the detector, and includes processes that were added in the final $\nu_e$ appearance analysis \cite{mb_lowe}. The $M_{A}^{QE}$ appearing in the nucleon axial vector form factor, and the Pauli blocking parameter, $\kappa$, used to parametrize neutrino quasi-elastic scattering on carbon, were adjusted by fits to MiniBooNE data, as were the coherent pion cross sections \cite{mb_numuccqe,mb_pi0}. The $M_{A}^{QE}$ and $\kappa$ values of $1.23\pm0.08$ GeV and $1.022\pm0.021$, respectively, were used in this analysis. Two additional parameters, $M^{QE,H}_A=1.13\pm0.10$ GeV and $M^{1\pi,H}_A=1.10\pm0.10$ GeV, were introduced in the analysis to parameterize antineutrino quasi-elastic scattering on hydrogen and single pion production on hydrogen. These processes have a non-negligible contribution in antineutrino running mode, where roughly 25\% of the antineutrino quasi-elastic scatters are on hydrogen rather than carbon.

The detector cannot differentiate (on an event-by-event basis) a $\nu_{\mu}$ from a $\bar{\nu}_{\mu}$ interaction, or a $\nu_e$ from a $\bar{\nu}_e$ interaction. Therefore, the reconstruction and selection requirements for $\bar{\nu}_e$-induced charged-current quasi-elastic (CCQE) events, which is the characteristic signature of any possible signal from $\bar{\nu}_{\mu}\rightarrow\bar{\nu}_e$ oscillations, are identical to those of the final neutrino mode analysis \cite{mb_lowe}.

To provide a constraint on $\bar{\nu}_e$ candidate events, a $\bar{\nu}_{\mu}$ CCQE sample is also formed by looking for events with a muon-like Cherenkov ring and a cluster of delayed PMT hits from the decay of the muon into an electron. The first cluster of PMT hits (muon subevent) is required to have more than 200 inner detector PMT hits, and no more than six outer (veto) PMT hits. A maximum of 200 inner detector and six veto PMT hits are required for the second subevent (decay electron), and a minimum time cut of 1000 ns between the first and second subevents is required to ensure PMT stability for proper charge response. After reconstruction, the first subevent vertex and the track end-point under the muon hypothesis are required to occur within the fiducial volume. The neutrino energy reconstructed from the outgoing muon energy and angle, $E^{QE}_{\nu}$, is required to satisfy $E_{\nu}^{QE}>150$ MeV. A cut on the separation distance between the muon and decay electron vertices as a function of reconstructed energy of the muon is also applied to provide rejection against backgrounds, mostly from CC $\pi^{+}$ interactions. For more details on the reconstruction method, see \cite{mb_recon}.

The oscillation parameters are extracted from a combined fit to $\bar{\nu}_e$ CCQE and $\bar{\nu}_{\mu}$ CCQE event distributions, following \cite{mb_lowe}. This fit method takes advantage of strong flux and cross section correlations among the $\bar{\nu}_e$ CCQE and $\bar{\nu}_{\mu}$ CCQE event samples, since any possible $\bar{\nu}_{\mu}\rightarrow\bar{\nu}_e$ signal, as well as some $\bar{\nu}_e$ backgrounds, interact through the same process as $\bar{\nu}_{\mu}$ CCQE events, and are related to $\bar{\nu}_{\mu}$ CCQE events through the same $\pi^+$ or $\pi^-$ decay chain at production. These correlations enter through the off-diagonal elements of the covariance matrix used in the $\chi^2$ calculation, relating the contents of the bins of the $\bar{\nu}_e$ CCQE and $\bar{\nu}_{\mu}$ CCQE distributions. This procedure maximizes the sensitivity to $\bar{\nu}_{\mu}\rightarrow\bar{\nu}_e$ oscillations when systematic uncertainties are included \cite{Schmitz:2008zz}.

A sample of 14,107 data events passing $\bar{\nu}_{\mu}$ CCQE selection requirements is used in the analysis. This sample is compared to a MonteCarlo prediction which has been corrected to match the observed $\bar{\nu}_{\mu}$ CCQE data through a normalization factor of 1.22 applied to events from $\pi^{-}$ decays in the beam, and 0.93 applied to events from $\pi^+$ decays in the beam. These normalization factors are extracted from a fit to the angular distributions of the outgoing $\mu^{+}$ and $\mu^{-}$ in $\bar{\nu}_{\mu}$ and $\nu_{\mu}$ CCQE interactions \cite{nubarLOI}. These two factors result in an overall 15\% normalization correction which is covered by flux and cross section uncertainties. The same normalization correction is also applied to all possible signal events which share the same parent ($\pi^{-}$) as $\bar{\nu}_{\mu}$ CCQE events. The normalization correction is accounted for in the oscillation fit by a reduction in the quoted effective degrees of freedom by one unit. After correction, the sample contains 95\% $\bar{\nu}_{\mu}$ and $\nu_{\mu}$ produced in pion decays and 2.4\% $\bar{\nu}_{\mu}$ and $\nu_{\mu}$ produced in kaon decays. The neutrino content of the sample is $22$\%. The majority of events (71\%) are true CCQE interactions, with CC$\pi^{\pm}$ interactions being the dominant source of background (20\%). This sample is included in the $\bar{\nu}_e$ appearance fits as a function of 8 bins of reconstructed neutrino energy, $E^{QE}_{\nu}$, ranging from 0 to 1900 MeV. 

\begin{table}[t]
\vspace{-0.1in}
\caption{\label{signal_bkgd} \em The expected number of events
for different $E_\nu^{QE}$ ranges (in MeV) from all of the backgrounds in the $\bar{\nu}_e$ appearance analysis and for the LSND central expectation (0.26\% oscillation probability) of $\bar{\nu}_{\mu}\rightarrow\bar{\nu}_e$ oscillations, for $3.39\times10^{20}$ POT.
}
\small
\begin{ruledtabular}
\begin{tabular}{cccc}
Process&$200-300$&$300-475$&$475-1250$\\
\hline
$\stackrel{\small{(-)}}{\nu_\mu}$ CCQE & 1.3 & 1.6 & 1.2 \\
NC $\pi^0$ & 14.4 & 10.2 & 7.2 \\
NC $\Delta \rightarrow N \gamma$ & 1.7 & 4.9 & 2.0\\
External Events & 2.2 & 2.5 & 1.9 \\
Other $\stackrel{\small{(-)}}{\nu_\mu}$ & 2.0 & 1.8 & 2.2 \\
\hline
$\stackrel{\small{(-)}}{\nu_e}$ from $\mu^{\pm}$ Decay & 2.3 & 5.9 & 17.1 \\
$\stackrel{\small{(-)}}{\nu_e}$ from $K^{\pm}$ Decay & 1.4 & 3.8 & 11.7 \\
$\stackrel{\small{(-)}}{\nu_e}$ from $K^0_L$ Decay & 0.8 & 2.4 & 13.1 \\
Other $\stackrel{\small{(-)}}{\nu_e}$ & 0.5 & 0.6 & 1.21 \\
\hline
Total Background &$26.7$&$33.6$&$57.8$ \\
0.26\% $\bar{\nu}_{\mu}\rightarrow\bar{\nu}_e$ & 0.6 & 3.7 & 12.6\\
\end{tabular}
\vspace{-0.2in}
\end{ruledtabular}
\normalsize
\end{table}

Table \ref{signal_bkgd} shows the number of predicted $\bar{\nu}_e$ CCQE background events for different ranges of $E_\nu^{QE}$. The background estimates include both antineutrino and neutrino events, the latter representing $\sim44$\% of the total. The predicted backgrounds to the $\bar{\nu}_e$ CCQE sample are constrained by internal measurements at MiniBooNE. These measurements use event samples from regions in reconstructed kinematic variables where any possible signal from $\bar{\nu}_{\mu}\rightarrow\bar{\nu}_e$ is negligible, in order to preserve blindness. The constrained backgrounds include NC $\pi^{0}$ events, $\Delta\rightarrow N\gamma$ radiative events, and events from interactions outside the detector. The NC $\pi^0$ background events are adjusted in bins of $\pi^0$ momentum according to a direct $\pi^0$ rate measurement in antineutrino mode, following \cite{mb_pi0}, which uses events reconstructed near the $\pi^{0}$ mass peak. The size of the applied correction to the total NC $\pi^{0}$ rate is less than $10$\%. The $\Delta\rightarrow N\gamma$ rate is indirectly constrained, being related to the measured $\pi^0$ rate through a branching fraction and final state interaction correction. The rate of backgrounds from external interactions is constrained through a direct measurement at MiniBooNE, using a separate event sample where the rate of external interaction events is enhanced. Other backgrounds from mis-identified $\nu_{\mu}$ or $\bar{\nu}_{\mu}$ receive the $\bar{\nu}_{\mu}$ CCQE normalization correction according to their parentage at production ($\pi^+$ or $\pi^-$). Intrinsic $\nu_e$ and $\bar{\nu}_e$ events from the $\pi\rightarrow\mu$ decay chain also receive this normalization.

Systematic uncertainties are determined by considering the effects on the $\bar{\nu}_{\mu}$ and $\bar{\nu}_e$ CCQE rate prediction of variations of fundamental parameters within their associated uncertainty. These include uncertainties on the flux estimate, including beam modeling and hadron production at the target, uncertainties on neutrino cross sections, most of which are determined by in-situ cross-section measurements at MiniBooNE or other experimental or theoretical sources, and uncertainties on detector modeling and reconstruction. By considering the variation from each source of systematic uncertainty on the $\bar{\nu}_e$ CCQE signal, background, and $\bar{\nu}_{\mu}$ CCQE prediction as a function of $E_{\nu}^{QE}$, a covariance matrix in bins of $E^{QE}_{\nu}$ is constructed, which includes correlations between $\bar{\nu}_e$ CCQE (signal and background) and $\bar{\nu}_{\mu}$ CCQE. This covariance matrix is used in the $\chi^2$ calculation of the oscillation fit.

\begin{figure}[tbp]
\vspace{-0.1in}
\centerline{\includegraphics[angle=-90, width=7.2cm, trim=30 130 60 80]{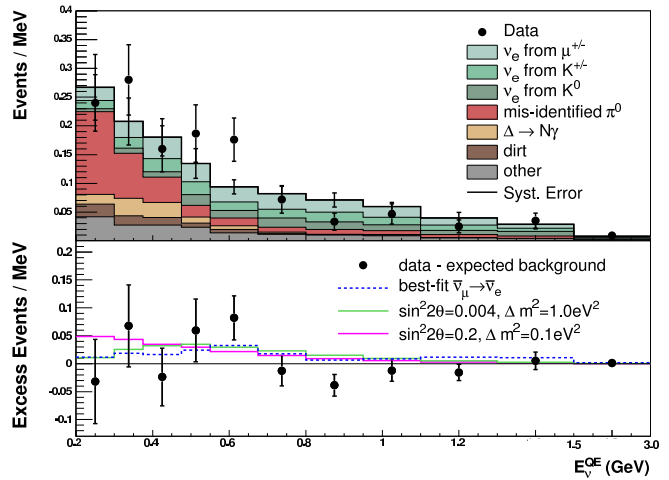}}
\vspace{0.1in}
\caption{Top: The $E_\nu^{QE}$ distribution for $\bar{\nu}_e$ CCQE data (points
 with statistical errors) and background (histogram with unconstrained systematic errors). Bottom: The event excess as a function of $E_\nu^{QE}$. Also shown are the expectations from the best oscillation fit and from neutrino oscillation parameters in the LSND allowed region. The error bars include both statistical and systematic errors.}
\label{data_mc1}
\vspace{-0.2in}
\end{figure}

\begin{table}[b]
\vspace{-0.2in}
\caption{\label{signal_bkgd3} \em The number of data, 
background, and excess events in the $\bar{\nu}_e$ analysis for different
$E_\nu^{QE}$ ranges. The corresponding numbers from the $\nu_e$ analysis \cite{mb_lowe} are on the right. The uncertainties include both statistical and constrained systematic errors.}
\begin{ruledtabular}
\begin{tabular}{c|cc}
Event Sample&$\bar{\nu}_e$ Analysis&$\nu_e$ Analysis\cite{mb_lowe} \\
&($3.39\times10^{20}$ POT)&($6.46\times10^{20}$ POT)\\
\hline
$200-475$ MeV&& \\
Data&61&544 \\
Background&$61.5 \pm 11.7$&$415.2 \pm 43.4$ \\
Excess&$-0.5 \pm 11.7$ ($-0.04 \sigma$)&$128.8 \pm 43.4$ ($3.0 \sigma$) \\
\hline
$475-1250$ MeV&& \\
Data&61&408 \\
Background&$57.8 \pm 10.0$&$385.9 \pm 35.7$ \\
Excess&$3.2 \pm 10.0$ ($0.3 \sigma$)&$22.1 \pm 35.7$ ($0.6 \sigma$) \\
\end{tabular}
\end{ruledtabular}
\end{table}

Figure \ref{data_mc1} (top) shows the $E_\nu^{QE}$ distribution for $\bar{\nu}_e$ CCQE observed data and background. A total of 144 events pass the $\bar{\nu}_e$ event selection requirements with $200<E_\nu^{QE}<3000$ MeV. The data agree with the background prediction within systematic and statistical uncertainties. Fig.~\ref{data_mc1} (bottom) shows the event excess as a function of $E_\nu^{QE}$. Also shown are expectations from the best $\bar{\nu}_{\mu}\rightarrow\bar{\nu}_e$ oscillation parameters returned by the fit and from two other sets of neutrino oscillation parameters from the LSND allowed region \cite{lsnd}. The best oscillation fit for $200<E_\nu^{QE}<3000$ MeV corresponds to ($\Delta m^2$, $\sin^22\theta$) $=$ (4.42 eV$^2$, 0.004), and has a $\chi^2$ of 18.2 for 16 degrees of freedom ($DF$), corresponding to a $\chi^2$-probability of 31\%. The null fit yields $\chi^2/DF=24.5/18$, with a $\chi^2$-probability of 14\%. A fit to $475<E_\nu^{QE}<3000$ MeV returns similar best-fit oscillation parameters, ($\Delta m^2$, $\sin^22\theta$) $=$ (4.42 eV$^2$, 0.005), with $\chi^2/DF=15.9/13$ and a $\chi^2$-probability of 25\%. The null fit to $475<E_\nu^{QE}<3000$ MeV yields $\chi^2/DF=22.2/15$, with a $\chi^2$-probability of 10\%. The number of data, background, and excess events for different $E_\nu^{QE}$ ranges are summarized in Table \ref{signal_bkgd3}. No significant event excess is observed for $E_\nu^{QE}>475$ MeV. Furthermore, no significant excess is observed for $E_\nu^{QE}<475$ MeV, to be compared to a $3.0 \sigma$ excess observed for $200<E_\nu^{QE}<475$ MeV in the $\nu_e$ appearance analysis \cite{mb_lowe}. 

\begin{figure}[tbp]
\vspace{-0.2in}
\centerline{\includegraphics[angle=-90,width=6cm, trim=140 160 120 130]{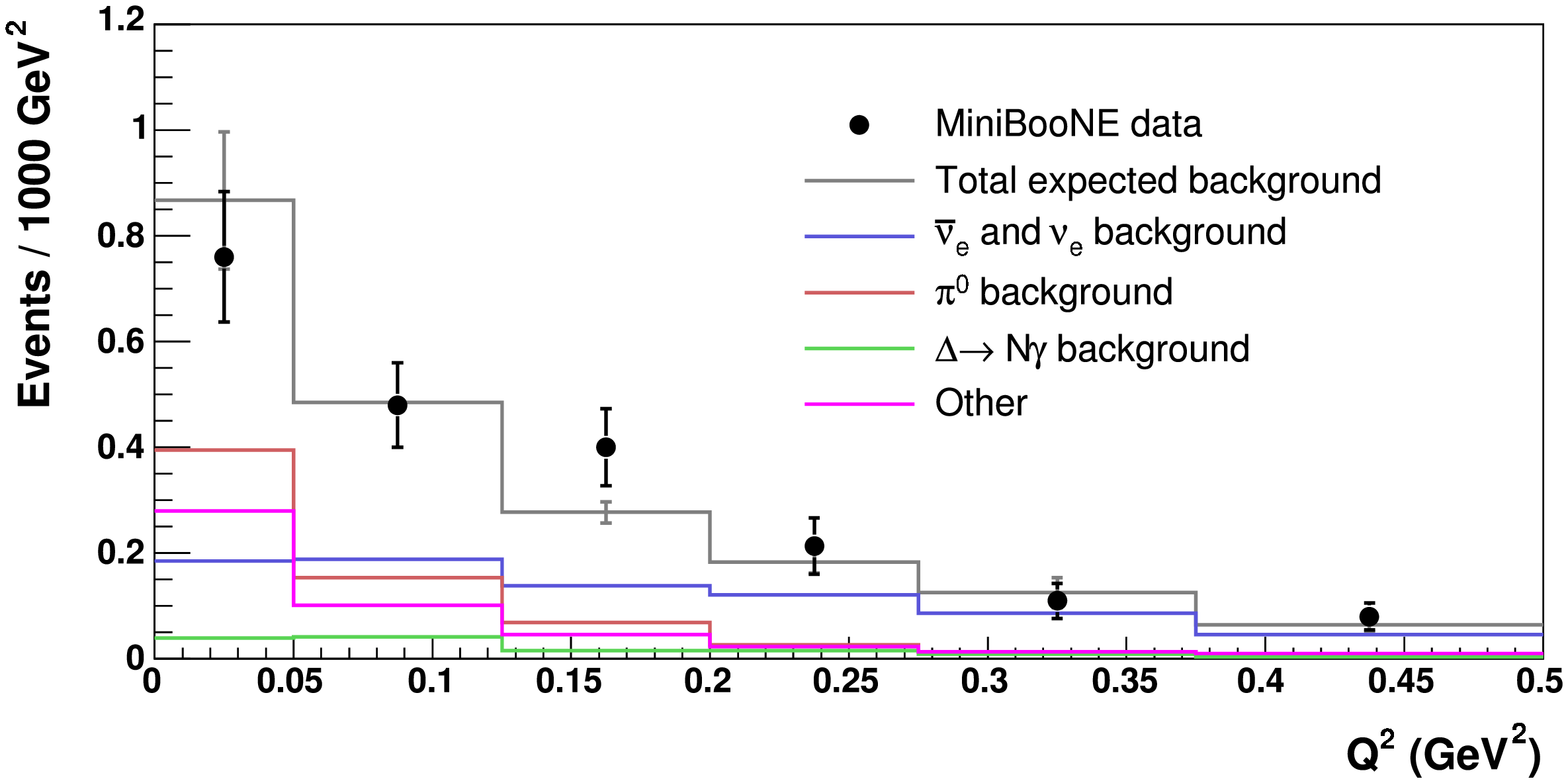}}
\centerline{\includegraphics[angle=-90,width=6cm, trim=110 160 80 130]{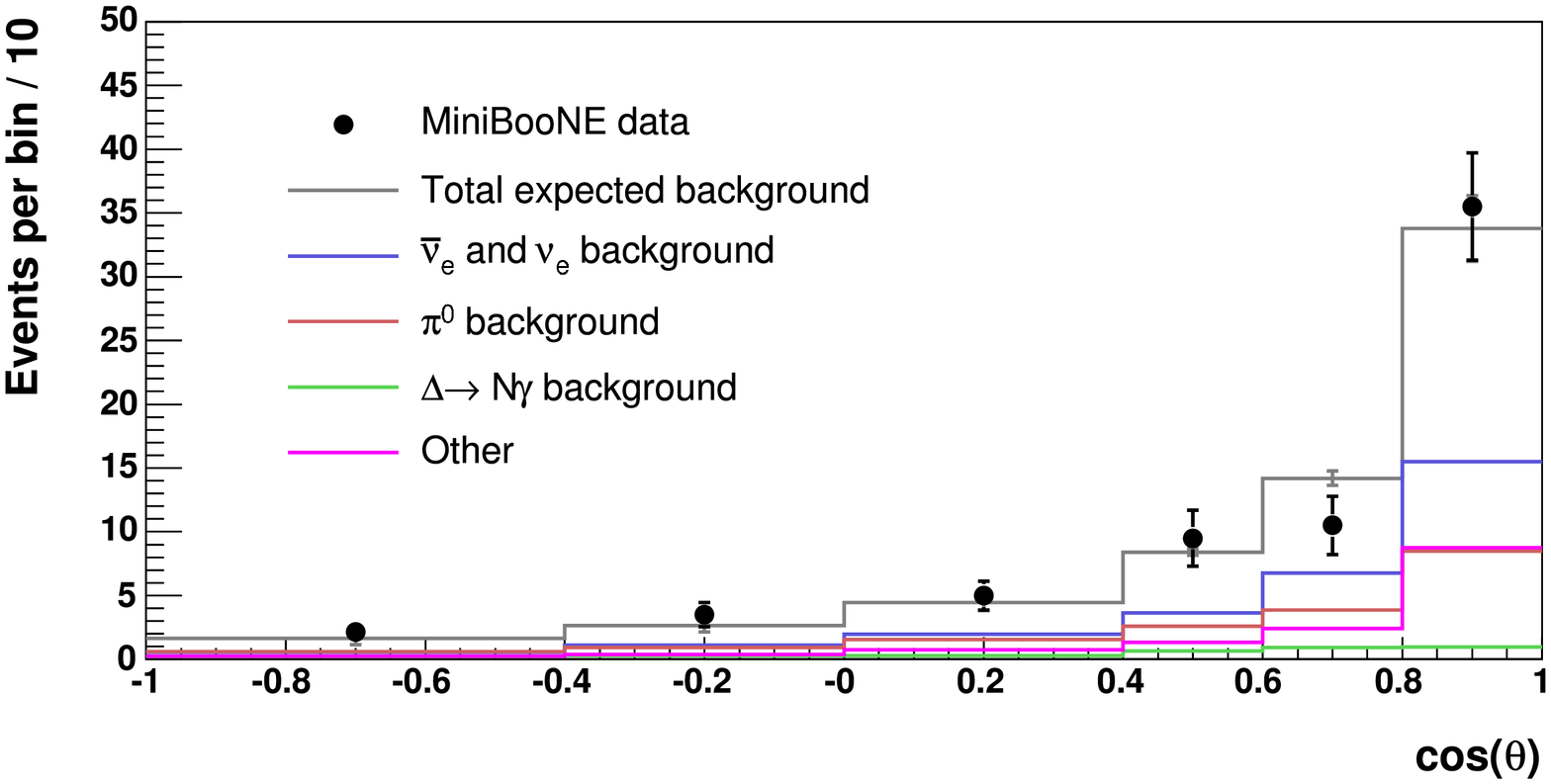}}
\vspace{0.1in}
\caption{The $Q^2$ (top panel) and $\cos(\theta)$ (bottom panel) distributions for data (points with statistical errors) and backgrounds (histogram with constrained systematic errors) for $E_\nu^{QE} > 200$ MeV. Also shown are the expected distributions from intrinsic $\bar{\nu}_e$ and $\nu_e$, and NC $\pi^0$ and $\Delta \rightarrow N \gamma$ backgrounds.} 
\label{data_mc4}
\vspace{-0.2in}
\end{figure}

The $\bar{\nu}_e$ data also exhibit reasonable agreement with predicted background in other reconstructed kinematic variables. Fig.~\ref{data_mc4} shows the observed and predicted event distributions as functions of reconstructed $Q^2$ and $\cos (\theta)$ for $200 < E_\nu^{QE} < 3000$ MeV. $Q^2$ is determined from the energy of the outgoing lepton and its scattering angle with respect to the incident neutrino direction ($\theta$) assuming CCQE scattering. Also shown in the figures are the predicted distributions from NC $\pi^0$ and $\Delta \rightarrow N \gamma$ backgrounds, which are events with a photon in the final state. The null $\chi^2$ values from these comparisons are both acceptable, at $\chi^2/DF=10.6/11$ and $\chi^2/DF=8.4/11$ for $Q^2$ and $\cos (\theta)$, respectively.

\begin{figure}[tbp]
\centerline{\includegraphics[height=4in,trim=100 50 100 30]{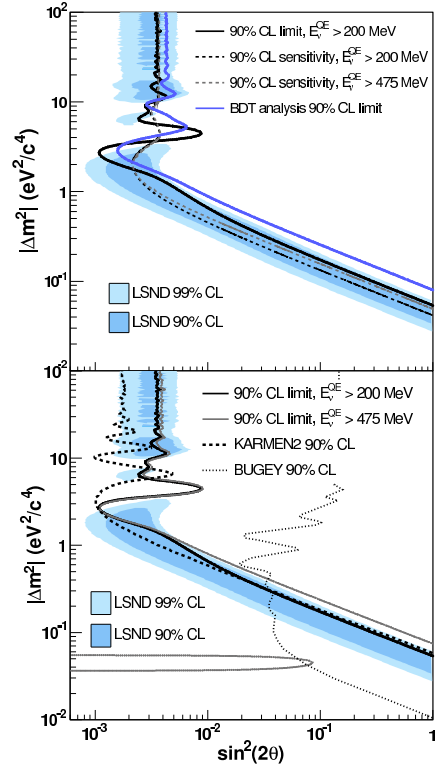}}
\caption{Top: MiniBooNE 90\% C.L. limit (solid black) and sensitivity (dashed black) for events with $E^{QE}_{\nu} > 200$ MeV, within a two neutrino $\bar{\nu}_{\mu}\rightarrow\bar{\nu}_e$ oscillation model. Also shown is the sensitivity for $E^{QE}_{\nu} > 475$ MeV (dashed grey), and the limit from the BDT analysis (solid blue) for $E^{QE}_{\nu} > 500$ MeV. Bottom: Limits from MiniBooNE for $E^{QE}_{\nu} > 200$ MeV and $E^{QE}_{\nu} > 475$ MeV, KARMEN \cite{karmen}, and Bugey \cite{bugey}. The island contour in the bottom left corner is a lower $\sin^2 2\theta$ limit from a fit to $E^{QE}_{\nu} > 475$ MeV, excluding the points left of the line at 90\% C.L.. The MiniBooNE and Bugey curves are 1-sided limits for $\sin^2 2\theta$ corresponding to $\Delta\chi^2 = 1.64$, while the KARMEN curve is a ``unified approach'' 2D contour. The shaded areas show the 90\% and 99\% C.L. LSND allowed regions.}
\label{limit}
\vspace{-0.2in}
\end{figure}

The absence of a significant excess allows MiniBooNE to place a limit on $\bar{\nu}_{\mu}\rightarrow\bar{\nu}_e$ oscillations as shown in Fig.~\ref{limit}. The bottom panel of the figure shows the MiniBooNE limits obtained from fits to events with $E^{QE}_{\nu} > 200$ MeV and $E^{QE}_{\nu} > 475$ MeV. Each 90\% C.L. limit on $\sin ^2 2\theta$ is obtained by a single-sided raster scan of the parameter space, where a $\Delta\chi^2=\chi^2_{limit}-\chi^2_{best\ fit}<1.64$ cut is applied for each slice in $\Delta m^2$. The two limits are in agreement, with the one obtained for $E^{QE}_{\nu} > 200$ MeV placing a stronger bound for low $\Delta m^{2}$ oscillations, due to its slightly better sensitivity in that region (see top panel of Fig.~\ref{limit}). At higher $\Delta m^{2}$ values, both limits approach the corresponding sensitivities of the experiment, but at lower $\Delta m^2$ both limits are noticeably worse due to the observed data fluctuation between $475<E_\nu^{QE}<675$ MeV. The significance of that fluctuation in the $475<E_\nu^{QE}<675$ MeV range is 2.8$\sigma$ (statistical $\oplus$ constrained systematic). 

Following \cite{mb_osc}, a secondary analysis based on Boosted Decision Trees (BDT) has been performed and used as a cross-check for the oscillation analysis in the energy region $E^{QE}_{\nu} > 500$ MeV, where the BDT analysis is not dominated by systematic uncertainties. No significant excess of events is observed with the BDT analysis, yielding the limit shown in the top panel of Fig.~\ref{limit}. Although the limit from the BDT analysis is not as stringent as the main result discussed above, the two analyses are complementary and yield consistent results.

In summary, MiniBooNE observes no significant excess of $\bar{\nu}_e$ events in the energy region $E_\nu^{QE} > 200$ MeV, for a data sample corresponding to $3.39\times10^{20}$ POT. Thus, with current statistics, MiniBooNE places a limit on two-neutrino $\bar{\nu}_{\mu}\rightarrow\bar{\nu}_e$ oscillations shown by the black line in Fig.~\ref{limit}. The result is inconclusive with respect to small amplitude mixing suggested by the LSND data, but more antineutrino data, which are currently being collected, will provide additional information. More constraints may also be provided by the off-axis NuMI beam data collected in MiniBooNE \cite{numi}. Interestingly, MiniBooNE observes no significant excess of $\bar{\nu}_e$ events in the low energy region $200 < E_\nu^{QE} < 475$ MeV. The absence of an excess at low energy in antineutrino mode should help distinguish between several hypotheses suggested as explanations for the low energy excess observed in neutrino mode. 


\begin{acknowledgments}
We acknowledge the support of Fermilab, the Department of Energy,
and the National Science Foundation, and
we acknowledge Los Alamos National Laboratory for LDRD funding. We also acknowledge the use of Condor software for the analysis of the data.
\end{acknowledgments}


\bibliography{prl}

\end{document}